\documentclass[12pt]{iopart}

\usepackage{graphicx}
\usepackage{iopams}
\usepackage[square,sort&compress,numbers]{natbib}
\begin{document}

\title[Coherently combining data for searches]{Coherently combining short data segments for all-sky semi-coherent continuous gravitational wave searches}
\author{E Goetz$^{1,2}$ and K Riles$^1$}
\address{$^1$ University of Michigan, Department of Physics, 450 Church St., Ann Arbor MI 48109, USA}
\address {$^2$ Albert-Einstein-Institut, Max-Planck-Institut f\"ur Gravitationsphysik, D-30167 Hannover, Germany }
\ead{egoetz@umich.edu}

\begin{abstract}
We present a method for coherently combining short data segments from gravitational-wave detectors to improve the sensitivity of semi-coherent searches for continuous gravitational waves. All-sky searches for continuous gravitational waves from unknown sources are computationally limited. The semi-coherent approach reduces the computational cost by dividing the entire observation timespan into short segments to be analyzed coherently, then combined together incoherently. Semi-coherent analyses that attempt to improve sensitivity by coherently combining data from multiple detectors face a computational challenge in accounting for uncertainties in signal parameters. In this article, we lay out a technique to meet this challenge using summed Fourier transform coefficients. Applying this technique to one all-sky search algorithm called TwoSpect, we confirm that the sensitivity of all-sky, semi-coherent searches can be improved by coherently combining the short data segments. For misaligned detectors, however, this improvement requires careful attention when marginalizing over unknown polarization parameters. In addition, care must be taken in correcting for differential detector velocity due to the Earth's rotation for high signal frequencies and widely separated detectors.
\end{abstract}


\maketitle

\section{Introduction}
All-sky searches for continuous gravitational waves are computationally limited because of the enormous parameter space that must be searched over. A common approach to reduce the computational cost divides the total observational timespan into shorter segments~\citep{S2Hough,StackSlideMethod,S4PSH,EarlyS5PowerFlux,FullS5PowerFlux,TwoSpectS6VSR23}. The short segments of calibrated strain data are individually Fourier transformed before being incoherently combined together---a method known as semi-coherent analysis. While this approach saves computational cost, it reduces the overall sensitivity of the searches to the continuous gravitational waves compared to what is theoretically possible with unlimited computing resources.

Various semi-coherent all-sky continuous gravitational-wave searches have been carried out on data collected with the Laser Interferometer Gravitational wave Observatory (LIGO)~\citep{LIGOifo2009} and Virgo~\citep{VirgoDetector} gravitational-wave interferometers. Several of those searches have coherently analyzed data among multiple detectors using the multi-detector $\mathcal{F}$-statistic analysis technique~\citep{S2cwSearch,S4EatH,EarlyS5EatH,FullS5EatH}. The $\mathcal{F}$-statistic approach coherently combines the data over time periods long enough to require coherent demodulation for Doppler effects from the Earth's motion and results in higher computational costs~\citep{JKS1998,MultiFstat,MultiFstatMetric}. Another coherent multi-detector approach, taken by the PowerFlux algorithm, explicitly searches over different values of the gravitational-wave polarization angle when coherently combining data from different detectors~\citep{FullS5PowerFlux}.

When a continuous gravitational wave signal due to a rapidly rotating neutron star is present in a short observation time ($\lesssim$30~min), the signal is nearly monochromatic in the detector rest frame. Hence, it can prove beneficial to combine coherently the Short Fourier Transformations (SFTs) taken at the same time from separate detectors before proceeding to compute the power spectrum (and any other data analysis product). We expect an improvement in sensitivity with respect to an incoherent combination due to better discrimination of signal versus noise. The degree of improvement depends on the specific construction of an algorithm's detection statistic.

Here, we explore the potential strain sensitivity gains from coherent summing of SFTs in the TwoSpect program~\citep{TwoSpectMethod,TwoSpectS6VSR23} developed for continuous gravitational wave searches in binary systems. This investigation assumes that detectors have equal strain noise values (e.g. $S_h^{1/2}=10^{-23}$~Hz$^{-1/2}$) and equal observation time spans. We find that for misaligned detectors, the sensitivity gain is sensitive to marginalization over unknown source parameters.

\section{Gravitational wave signal}
Consider the phase of a continuous gravitational wave signal in a short coherent interval, $T_{\rm{SFT}} = 1800$~s or less, with frequency $f$ and initial phase $\phi_0$ in the Solar System barycenter (SSB),
\begin{equation}
\Phi(\tau) = 2\pi f \tau(t)+\phi_0\,,
\end{equation}
where the time in the SSB, $\tau$, is given in terms of the time in the detector frame, $t$. Here, it is assumed the intrinsic frequency evolution of the source is very small compared to the SFT coherence time (i.e. $f^{(n)}\ll T_{\rm{SFT}}^{-(n+1)}$), and frequency modulation due to a source in a binary orbit is small over the timescale of $T_{\rm{SFT}}$. Of particular focus here in this work are coherent intervals taken at the same time in different detectors. This is because the binary phase evolution is unknown in an all-sky, unknown binary search, making assumptions of known coherence over extended intervals questionable.

The phase can be written then in terms of the detector time
\begin{equation}\label{eq:sigphasedetectortime}
\Phi^X(t) = 2\pi f\left(t+\frac{\vec{r}^X(t)\cdot\vec{n}}{c}\right)+\phi_0\,,
\end{equation}
where the detector is denoted by the index $X$, $\vec{r}^X$ is the vector pointing from the SSB to the detector, $\vec{n}$ is the vector that points from the SSB to the sky location, and $c$ is the speed of light.

Assuming the time of coherent observation is short enough such that the phase of the signal is approximately linear, equation~(\ref{eq:sigphasedetectortime}) can be Taylor-expanded around the midpoint time, $t_m$, of the coherent observation. The phase is then given by
\begin{equation}
\fl \Phi^X(t) \simeq \phi_0 + 2\pi ft_m + 2\pi f(t-t_m) + 2\pi f\left[\frac{\vec{r}^X(t_m)\cdot\vec{n}}{c}+\frac{\vec{v}^X(t_m)\cdot\vec{n}}{c}(t-t_m)\right]\,.
\end{equation}
Using $\hat{\phi}_0 \equiv \phi_0+2\pi ft_m$, $\Delta t^X_{\rm{SSB}} \equiv \vec{r}^X(t_m)\cdot\vec{n}/c$, and $\dot{t}^X_{\rm{SSB}} \equiv 1+\vec{v}^X(t_m)\cdot\vec{n}/c$, then $\Phi^X(t)$ can be written more simply as
\begin{equation}
\Phi^X(t) = \hat{\phi}_0+2\pi f\Delta t^X_{\rm{SSB}}+2\pi f\dot{t}^X_{\rm{SSB}}(t-t_m)\,.
\end{equation}

\subsection{Gravitational wave signal measured by a detector}
The signal strain measurable by a gravitational wave detector $X$ is given by
\begin{equation}
h^X(t) = h_0F^X_+(t)\frac{1+\cos^2\iota}{2}\cos\Phi^X(t)+h_0F^X_\times(t)\cos\iota\sin\Phi^X(t)\,,
\end{equation}
where $h_0$ is the amplitude of the gravitational wave, $\iota$ is the inclination angle of the source, and the detector antenna patterns, $F^X_+(t)$ and $F^X_\times(t)$, are given by~\citep{JKS1998}
\begin{eqnarray}
F^X_+(t) & = & a^X(t)\cos2\psi+b^X(t)\sin2\psi \\
F^X_\times(t) & = & b^X(t)\cos2\psi-a^X(t)\sin2\psi\,.
\end{eqnarray}
Here, $\psi$ is the wave polarization angle, and $a$ and $b$ are detector specific functions that depend on the location of the source on the sky (right ascension $\alpha$ and declination $\delta$) and the orientation and location of the detector on Earth. The detector antenna patterns are taken to be constant during an SFT because they are only slowly varying functions. The functions are evaluated at the midpoint of the SFT, $F^X_+(t_m)$ and $F^X_\times(t_m)$. Then, simplifying,
\begin{equation}\label{eq:hoft}
h^X(t) = \mathcal{A}_+^X\cos\Phi^X(t) + \mathcal{A}_\times^X\sin\Phi^X(t)\,,
\end{equation}
where
\begin{eqnarray}
\mathcal{A}_+^X & \equiv & h_0F^X_+(t_m)\frac{1+\cos^2\iota}{2} \\
\mathcal{A}_\times^X & \equiv & h_0F^X_\times(t_m)\cos\iota\,.
\end{eqnarray}

\subsection{Discrete Fourier transformation}
Semi-coherent analysis techniques rely on the Fourier transformation of the short segments of data. To combine different detectors' data, the Fourier coefficients must be understood for a signal. Equation~(\ref{eq:hoft}) can be written as
\begin{eqnarray}
h^X(t) & = & \frac{\mathcal{A}_+^X}{2}\left[{\rm e}^{{\rm i}\Phi^X(t)}+{\rm e}^{-{\rm i}\Phi^X(t)}\right] + \frac{\mathcal{A}_\times^X}{2{\rm i}}\left[{\rm e}^{{\rm i}\Phi^X(t)}-{\rm e}^{-{\rm i}\Phi^X(t)}\right] \\
& = & A^X{\rm e}^{{\rm i}\Phi^{\it X}(t)} + A^{{\it X} \ast}{\rm e}^{-{\rm i}\Phi^{\it X}(t)}\,,
\end{eqnarray}
where $A^X\equiv (\mathcal{A}_+^X-{\rm i}\mathcal{A}_\times^X)/2$, and $\ast$ indicates the complex conjugate. Then the sampled time series is given by
\begin{eqnarray}
h_j^X & = & A^X{\rm e}^{{\rm i}[\hat{\phi}_0+2\pi f\Delta t^X_{\rm{SSB}}+2\pi f\dot{t}^X_{\rm{SSB}}(j\Delta t-t_m)]} + {\rm c.c.} \\
& = & A^X{\rm e}^{{\rm i}\hat{\varphi}^X+2\pi {\rm i} f\dot{t}^X_{\rm{SSB}}T_{\rm{SFT}}j/N} + \rm{c.c.} \,,
\end{eqnarray}
where $j=0,1,\dots,N-1$, $\hat{\varphi}^X \equiv \hat{\phi}_0+2\pi f(\Delta t^X_{\rm{SSB}} -\dot{t}^X_{\rm{SSB}}t_m)$, $\Delta t = T_{\rm{SFT}}/N$ is the sample spacing, and ``c.c.'' is the complex conjugate.

The discrete Fourier transform is defined as
\begin{eqnarray}
\fl \tilde{h}_k^X & = & \Delta t\sum_{j=0}^{N-1} h_j^X{\rm e}^{-2\pi {\rm i} jk/N} \\*
\fl & = & \frac{T_{\rm{SFT}}}{N}\sum_{j=0}^{N-1}A^X{\rm e}^{{\rm i}\hat{\varphi}^X}e^{2\pi {\rm i}(f\dot{t}^X_{\rm{SSB}}T_{\rm{SFT}}-k)j/N} + A^{X\ast}{\rm e}^{-{\rm i}\hat{\varphi}^X}{\rm e}^{-2\pi {\rm i}(f\dot{t}^X_{\rm{SSB}}T_{\rm{SFT}}+k)j/N} \,,
\end{eqnarray}
where $k=0,1,2\dots,N-1$. The second term in the sum can be ignored because only positive frequencies are considered. This means the Fourier transform can be written
\begin{equation}
\tilde{h}_k^X = A^X{\rm e}^{{\rm i}\hat{\varphi}^X}\frac{T_{\rm{SFT}}}{N}\sum_{j=0}^{N-1}\left[{\rm e}^{2\pi {\rm i}(f\dot{t}^X_{\rm{SSB}}T_{\rm{SFT}}-k)/N}\right]^j\,.
\end{equation}
Observe that the sum is simply a geometric series.

Therefore, the Fourier transform is given by
\begin{equation}
\tilde{h}_k^X = A^X{\rm e}^{{\rm i}\hat{\varphi}^X}\frac{T_{\rm{SFT}}}{N}\frac{{\rm e}^{2\pi {\rm i}(f\dot{t}^X_{\rm{SSB}}T_{\rm{SFT}}-k)}-1}{{\rm e}^{2\pi {\rm i}(f\dot{t}^X_{\rm{SSB}}T_{\rm{SFT}}-k)/N}-1}\,,
\end{equation}
and in the large-$N$ limit, when there are a large number of samples for the SFT, this becomes
\begin{equation}
\tilde{h}_k^X \simeq A^X{\rm e}^{{\rm i}\hat{\varphi}^X}T_{\rm{SFT}}\frac{{\rm e}^{2\pi {\rm i}(f\dot{t}^X_{\rm{SSB}}T_{\rm{SFT}}-k)}-1}{2\pi {\rm i}(f\dot{t}^X_{\rm{SSB}}T_{\rm{SFT}}-k)}\,.
\end{equation}
The last factor is the large-$N$ limit of the Dirichlet kernel:
\begin{equation}
D(\delta_k^X)\equiv\frac{\exp\left(2\pi {\rm i}\delta_k^X\right)-1}{2\pi {\rm i}\delta_k^X}\,,
\end{equation}
where $\delta_k^X \equiv f\dot{t}_{\rm{SSB}}^XT_{\rm{SFT}}-k=(\hat{f}^X-f_k)/\Delta f$. The final result is thus,
\begin{equation}\label{eq:fouriercoeff}
\tilde{h}_k^X \simeq A^X{\rm e}^{{\rm i}\hat{\varphi}^X}T_{\rm{SFT}}D\left(\delta_k^X\right)\,.
\end{equation}

\subsection{Discrete Fourier transform with windowing}
It is often useful to apply a windowing function to the time-series data in order to suppress leakage of signal power into neighboring frequency bins. A discrete Fourier transform with windowing can be defined as
\begin{equation}
\tilde{h}_k = \frac{\Delta t}{C}\sum_{j=0}^{N-1} w_j h_j {\rm e}^{-2\pi {\rm i} jk/N}\,,
\end{equation}
where $C\equiv(\sum_{j=0}^{N-1}w_j^2/N)^{1/2}$. A common choice is the Hann window ($w_j=0.5\{1-\cos[2\pi j/(N-1)]\}$, $C=\sqrt{3/8}$), which provides good leakage suppression and is conveniently described in the Fourier domain. For a Hann-windowed SFT, equation~(\ref{eq:fouriercoeff}) takes a slightly modified form compared to the unwindowed case:
\begin{equation}
\tilde{h}_k^X = A^X{\rm e}^{{\rm i}\hat{\varphi}^X}\frac{T_{\rm{SFT}}}{C}D_h\left(\delta_k^X\right)\,,
\end{equation}
where the final term is defined as
\begin{equation}
D_h\left(\delta_k^X\right) \equiv \frac{D\left(\delta_k^X\right)}{2}-\frac{D\left(\delta_k^X+1\right)}{4}-\frac{D\left(\delta_k^X-1\right)}{4}\,.
\end{equation}
This can be further simplified to
\begin{equation}
D_h\left(\delta_k^X\right) = \frac{{\rm i}{\rm e}^{2\pi {\rm i}\delta_k^X}-{\rm i}}{4\pi\delta_k^X\left[\left(\delta_k^X\right)^2-1\right]}\,.
\end{equation}

\section{\label{sec:coherentSFTaddition}Coherent SFT addition}
An optimal approach to determine whether or not a signal is present in the data begins by computing the inner product of the nominal signal, $\bi{h}$, with the data, $\bi{x}$,
\begin{equation}
\langle\bi{h}|\bi{x}\rangle_k = \sum_{X=0}^{N-1}\tilde{h}_k^{X\ast}\tilde{x}_k^X\,,
\end{equation}
where $X$ runs over the $N$ different detectors, $\tilde{x}_k^X$ are the SFT coefficients in bin $k$ of detector $X$, and the nominal signal template $\bi{h}$ is independent of $h_0$. Suppose that the assumed signal for $X=0$ is factored out of the sum; then the inner product becomes
\begin{equation}\label{eq:coherentsum}
\langle\bi{h}|\bi{x}\rangle_k = \tilde{h}_k^{0\ast}\sum_{X=0}^{N-1}\frac{\tilde{h}_k^{X\ast}}{\tilde{h}_k^{0\ast}}\tilde{x}_k^X\,,
\end{equation}
and an important quantity to be computed is $\tilde{h}_k^{X\ast}/\tilde{h}_k^{0\ast}$.

To compute the ratio, we start by replacing $A^X$ in equation~(\ref{eq:fouriercoeff}),
\begin{equation}
\tilde{h}_k^{X\ast} = \left(\frac{\mathcal{A}_+^X+{\rm i}\mathcal{A}_\times^X}{2}\right){\rm e}^{-{\rm i}\hat{\varphi}^X}T_{\rm{SFT}}D^\ast\left(\delta_k^X\right)\,.
\end{equation}
Then, the ratio of Fourier coefficients from detector $X$ to those from detector 0 is
\begin{eqnarray}
\frac{\tilde{h}_k^{X\ast}}{\tilde{h}_k^{0\ast}} & = & \frac{\left(\mathcal{A}_+^X+{\rm i}\mathcal{A}_\times^X\right) {\rm e}^{-{\rm i}\hat{\varphi}^X} D^\ast\left(\delta_k^X\right)}{\left(\mathcal{A}_+^0+{\rm i}\mathcal{A}_\times^0\right) {\rm e}^{-{\rm i}\hat{\varphi}^0} D^\ast\left(\delta_k^0\right)} \\*
& = & \frac{\mathcal{A}_+^X+{\rm i}\mathcal{A}_\times^X}{\mathcal{A}_+^0+{\rm i}\mathcal{A}_\times^0} {\rm e}^{-2\pi {\rm i} f\tau} \frac{D^\ast\left(\delta_k^X\right)}{D^\ast\left(\delta_k^0\right)} \,,
\end{eqnarray}
where $\tau \equiv \Delta t^X_{\rm{SSB}} -\dot{t}^X_{\rm{SSB}}t_m - \Delta t^0_{\rm{SSB}} +\dot{t}^0_{\rm{SSB}}t_m$. Writing the first and last complex valued fractions in terms of magnitude and phase,
\begin{equation}
\frac{\tilde{h}_k^{X\ast}}{\tilde{h}_k^{0\ast}} = M_S{\rm e}^{{\rm i}\phi_S} {\rm e}^{-2\pi {\rm i} f\tau} M_D{\rm e}^{{\rm i}\phi_D} \,,
\end{equation}
where $M_S$ and $M_D$ are the magnitudes of the first and last fractions, and $\phi_S$ and $\phi_D$ are the phases of the first and last fractions, accounting for detector-signal phase parameters and detector motion, respectively.

When the SFTs use Hann windowing, the ratio of Fourier coefficients is nearly the same:
\begin{eqnarray}
\frac{\tilde{h}_k^{X\ast}}{\tilde{h}_k^{0\ast}} & = & M_S{\rm e}^{{\rm i}\phi_S} {\rm e}^{-2\pi {\rm i} f\tau}\frac{D_h^\ast\left(\delta_k^X\right)}{D_h^\ast\left(\delta_k^0\right)} \\*
& = & M_S{\rm e}^{{\rm i}\phi_S} {\rm e}^{-2\pi {\rm i} f\tau} M_{D,h}{\rm e}^{{\rm i}\phi_{D,h}} \label{eq:convertcoeffs}
\end{eqnarray}

Determining the values for the terms in equation~\ref{eq:convertcoeffs} is discussed in detail in section~\ref{sec:correctionFactor}. Note that coherence between detectors is assumed only on timescales of a single SFT, and equation~(\ref{eq:convertcoeffs}) is independent of the intrinsic gravitational wave amplitude, $h_0$.

\section{Coherent SFT analysis: an application using the TwoSpect algorithm}
The TwoSpect analysis algorithm~\citep{TwoSpectMethod} is designed to detect continuous gravitational wave signals from sources in binary systems. Previous results using this method were restricted to analyzing detector data separately~\citep{TwoSpectS6VSR23}. Employing equation~(\ref{eq:convertcoeffs}), we have implemented a coherent SFT addition stage at the beginning of the pipeline to increase the sensitivity of the algorithm and to demonstrate this methodology's effectiveness.

\subsection{Implementation}
The TwoSpect analysis relies on the periodic Doppler shift caused by the motion of the continuous gravitational wave source in the binary system. A detailed description of the data preparation is found in~\citep{TwoSpectMethod}; a summary is given here. The pipeline first computes SFTs of detector $h(t)$ data segments, typically 30 minutes in length. (It is after this step that the coherent sum of SFTs from different detectors is applied.) The magnitude squared (the ``power'') of the Fourier coefficients is computed for each frequency bin of all SFTs. The running mean over frequency is computed and subtracted for every individual SFT\footnote{In practice, the running median is computed so that sharp spectral features or potential gravitational wave signals do not distort the computation. The running mean is computed from the running median with a simple scale factor for an exponential distribution~\citep{S4PSH,TwoSpectMethod}}. The SFTs are then weighted by the appropriate detector antenna pattern for the given sky location and noise variance. In this way, SFTs that have lower noise and higher sensitivity for a given sky point are given greater weight than those with higher noise or reduced sensitivity. Then the SFTs are adjusted to account for each detector's motion, using the known detector velocity via ephemeris files. At this point, the only remaining Doppler shift of the gravitational wave signal is due to the motion of the source. Fourier transforms are computed for every frequency bin $k$ in the constructed noise- and antenna pattern-weighted periodogram.

The above description of SFT data preparation for the second Fourier transform is mathematically expressed as~\citep{TwoSpectMethod}
\begin{equation}\label{eq:TwoSpectTSofPowers}
\widetilde{P}_{ki} = \frac{F_i^2(P_{ki} - \langle P_k\rangle_i)}{(\langle P_k\rangle_i)^2}\left[\sum_{i^\prime=1}^M\frac{F_{i^\prime}^4}{(\langle P_k\rangle_{i^\prime})^2}\right]^{-1}\,,
\end{equation}
where $i=1,2\ldots,M$ are the SFT index values, $k$ is the SFT frequency bin index, $F_i$ is the detector antenna pattern (typically chosen to be $F_i^2 = F_{i+}^2 + F_{i\times}^2$), $P_k\equiv2|\tilde{x}_k|^2/T_{\rm{SFT}}$ is the Fourier ``power'', and $\langle\,\rangle$ indicates the running mean computed over the inner index and is a measure of the noise, $\langle P_k\rangle_i \approx S_{ki}$. The power in Gaussian noise-only data is exponentially distributed, so the mean is also equal to the standard deviation of the data. Equation~(\ref{eq:TwoSpectTSofPowers}) is appropriately normalized so that the units of $\widetilde{P}$ are strain squared ($h^2$). The coherent SFT sum directly affects $P_{ki}$, so let us determine this role.

The noise weighting of equation~(\ref{eq:TwoSpectTSofPowers}) can be included in equation~(\ref{eq:coherentsum}) as follows:
\begin{equation}
\langle\bi{h}|\bi{x}\rangle_k \rightarrow \langle\bi{h}|\bi{x}\bi{S}^{-1}\rangle_k\,,
\end{equation}
so that the power is given by
\begin{equation}
\frac{2\left|\langle\bi{h}|\bi{x}\bi{S}^{-1}\rangle_k\right|^2}{T_{\rm{SFT}}} = \frac{|\tilde{h}_{ki}^{0}|^2}{(S_{ki}^0)^2} \frac{2}{T_{\rm{SFT}}} \left|\tilde{x}_{ki}^0 + \sum_{X=1}^{N-1}\frac{\tilde{h}_{ki}^{X\ast} S_{ki}^0\tilde{x}_{ki}^X}{\tilde{h}_{ki}^{0\ast} S_{ki}^X}\right|^2 \equiv \frac{|\tilde{h}_{ki}^{0}|^2}{(S_{ki}^0)^2}\hat{P}_{ki}\,.
\end{equation}
In noise alone, $\tilde{x}_{ki}^X = \tilde{n}_{ki}^X$, the expectation value of the power is
\begin{equation}\label{eq:noiseCohSum}
\textrm{E}\left[\frac{2\left|\langle\bi{h}|\bi{n}\bi{S}^{-1}\rangle_k\right|^2}{T_{\rm{SFT}}}\right] = \frac{|\tilde{h}_{ki}^{0}|^2 S_{ki}^0}{(S_{ki}^0)^2}\left[1+\sum_{X=1}^{N-1}\frac{|\tilde{h}_{ki}^X|^2 S_{ki}^0}{|\tilde{h}_{ki}^0|^2 S_{ki}^X}\right] \equiv \frac{|\tilde{h}_{ki}^{0}|^2 S_{ki}^0}{(S_{ki}^0)^2}\mathcal{C}_{ki}\,,
\end{equation}
where the cross-terms of $\tilde{n}_{ki}^X$ average to zero and $\mathcal{C}_{ki}$ is real-valued. In a Fourier bin dominated by a gravitational wave signal, $\tilde{x}_{ki}^X = h_0\tilde{h}_{ki}^X$ the power of the coherent sum is
\begin{equation}
\frac{2\left|\langle\bi{h}|h_0\bi{h}\bi{S}^{-1}\rangle_k\right|^2}{T_{\rm{SFT}}} = \frac{2h_0^2|\tilde{h}_{ki}^{0}|^4}{T_{\rm{SFT}}(S_{ki}^0)^2}\left|1+\sum_{X=1}^{N-1}\frac{|\tilde{h}_{ki}^X|^2 S_{ki}^0}{|\tilde{h}_{ki}^0|^2 S_{ki}^X}\right|^2 \equiv \frac{2h_0^2|\tilde{h}_{ki}^{0}|^4}{T_{\rm{SFT}}(S_{ki}^0)^2}\mathcal{C}_{ki}^2\,,
\end{equation}
where it is assumed that the values of $\bi{S}$ are taken from the pure noise case. Observe that for a bin-centered signal ($\delta_k=0$) and circularly polarized gravitational waves ($\cos\iota=\pm 1$), then $|\tilde{h}_{ki}^0|^2 = T_{\rm{SFT}}^2F_i^2/4$. Thus, we establish that coherently summing SFTs changes equation~(\ref{eq:TwoSpectTSofPowers}) as
\begin{equation}
\widetilde{P}_{ki} = \frac{F_i^2(\hat{P}_{ki} - \mathcal{C}_{ki}\langle P_k\rangle_i)}{(\langle P_k\rangle_i)^2}\left[\sum_{i^\prime=1}^M\frac{F_{i^\prime}^4\mathcal{C}_{ki^\prime}^2}{(\langle P_k\rangle_{i^\prime})^2}\right]^{-1}\,,
\end{equation}
where $F_i$ and $\langle P_k\rangle_i$ remain computed over detector $X=0$\footnote{The factor of $2/T_{\rm{SFT}}$ (not shown) is simply a normalization factor accounted for in the analysis.}. Therefore, the coherent sum of SFTs can be performed in the first stage of the pipeline before the power is computed with minimal reorganization of the TwoSpect analysis pipeline.

\subsection{\label{sec:correctionFactor}Determining the SFT correction factor}
To coherently add SFTs, the values of $M_S$, $\phi_S$, $f\tau$, $M_{D,h}$ and $\phi_{D,h}$ must, in principle, be known. If computational cost were not an issue, these values could be searched over along with other signal parameters. For realistic searches, however, a tradeoff in sensitivity is made to reduce the computational burden. The value of $\tau$ is known from ephemeris values of the Earth and Sun for each point on the sky. For the other terms, assumptions must be made to reduce the computational burden, as discussed below.

The first assumption is that $f\approx f_k$, meaning that the putative signal frequency is the bin-centered frequency of each frequency bin $f_k$. This assumption can cause a somewhat larger phase error at higher frequencies because the frequency bin that a true signal would be located in will have an incorrect phase correction applied. This error can be mitigated, however, by explicitly searching over specific values of $f$, although this approach substantially increases the computational cost.

A second assumption is that $M_{D,h}\approx1$, meaning that the Hann-windowed Dirichlet kernel magnitudes are approximately equal in each detector. Estimated values for $\delta_k^0$ and $\delta_k^X$ can be computed using $f\approx f_k$. Importantly, especially for high frequencies and widely separated detectors, SFT coefficients of detector $X$ must be corrected due to the relative detector velocities of detector $X$ and detector 0 with respect to a given sky location. The approximate correction is made by shifting SFT coefficients by integer frequency bins. The estimated number of frequency bins to shift is given by
\begin{equation}
\Delta\delta_k = \textrm{round}(\delta_k^0) - \textrm{round}(\delta_k^X)\,,
\end{equation}
so that $\delta_k^X\rightarrow\delta_k^X+\Delta\delta_k$. This effectively shifts the SFT coefficients of detector $X$ by $\Delta\delta_k$ frequency bins so that a putative signal would appear in the same SFT frequency bin of detector 0, helping to satisfy $M_{D,h}\approx1$. The estimated value for $\phi_{D,h}$ is calculated from $\delta_k^0$ and corrected $\delta_k^X$ values. An additional phase shift of $\pi$ is also added to $\phi_{D,h}$ when $|\textrm{floor}(\delta_k^0)-\textrm{floor}(\delta_k^X)|\geq1$ because there is a phase difference of $\pi$ radians in the Hann-windowed Dirichlet kernel for signals on different sides of a frequency bin.

Estimated values for $M_S$ and $\phi_S$ are perhaps the most difficult to obtain due to the complicated nature of how a continuous gravitational wave signal appears in an interferometer output. Left- and right-polarized gravitational waves are $\pi$ radians out of phase. Two possible approaches are considered here: 1) marginalizing over the full range of $\cos\iota$ and $\psi$ or 2) separately marginalizing over left- and right-polarized waves in $\cos\iota$ and the full range of $\psi$. When detectors are favorably aligned on the Earth (as the LIGO Hanford and Livingston detectors are constructed), then the first approach is a reasonable one. For poorly aligned detectors (as Virgo is poorly aligned to either LIGO Hanford or Livingston), then the second approach must be used in order to obtain reliable phase corrections for the SFT coefficients. The second approach, however, requires approximately twice the computing resources as the first since both left- and right-polarized waves must be considered separately.

\begin{figure}
\centering
\includegraphics[width=0.95\textwidth]{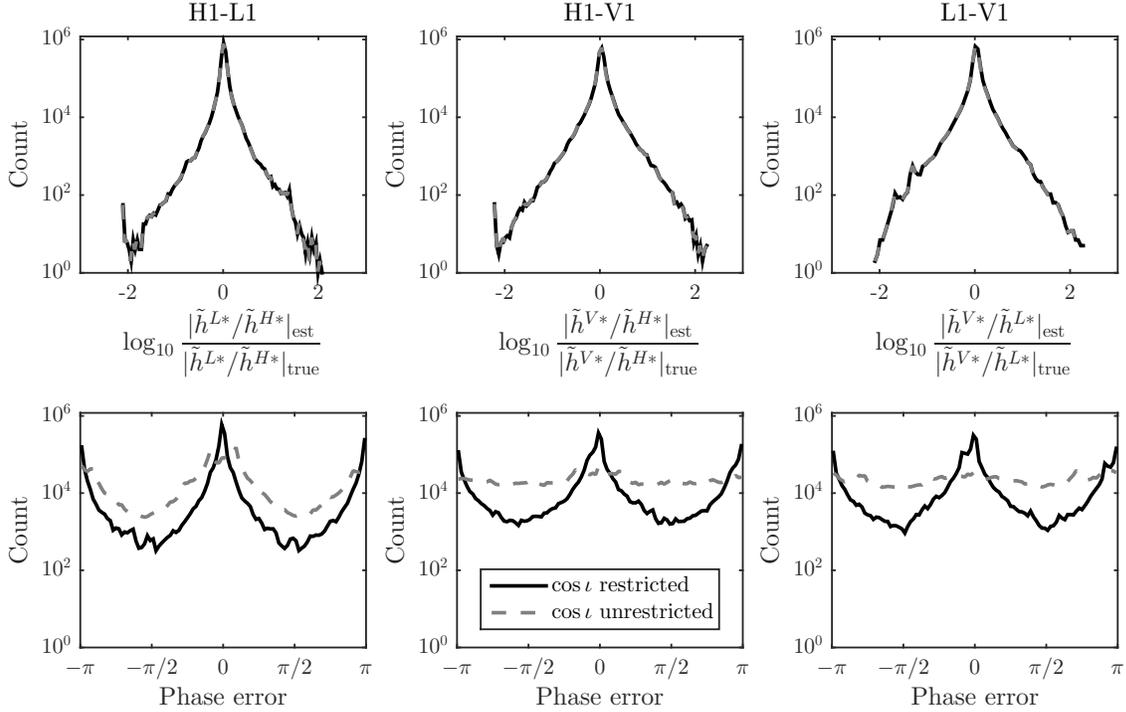}
\caption{\label{fig:coherentSFTerror}Histograms of the magnitude ratios and phase differences of estimated versus true SFT correction factors for signal frequencies centered at 1~kHz for three interferometer pairs and two different cases: $\cos\iota$ restricted marginalization (solid lines), or unrestricted marginalization (dashed lines). The phase errors of the restricted values are peaked around 0 and with small deviation from 0, whereas the unrestricted phase errors are more uniformly distributed between $-\pi$ and $\pi$, especially for H1-V1 and L1-V1 pairs.}
\end{figure}

Figure~\ref{fig:coherentSFTerror} shows the overall magnitude and phase errors of the corrections using these assumptions for three different pairs of interferometers, H1-L1, H1-V1, and L1-V1; SFT coherence time $T_{\rm{SFT}}=1800$~s; and using a representative high frequency band centered at 1~kHz. To construct the histograms, 500 simulations were tested, each with 4500 SFTs, with each simulation at a randomly chosen sky location, signal frequency $f=1$~kHz~$\pm0.277$~mHz, $\cos\iota\in[-1,1]$, $\psi\in[0,\pi)$ (all uniformly distributed). Given the known sky location, one can compute the true correction factors for the SFTs and compare them to the estimated correction factors. It is clear from the phase discrepancies that the non-aligned H1-V1, and L1-V1 pairs perform poorly compared to the H1-L1 pair in the full marginalization over $\cos\iota$. On the other hand, if doubling the computational cost can be tolerated, then separately marginalizing over left- and right-polarized gravitational waves allows for improvements for all three pairs. Still finer searching in multiple subsets of $\cos\iota$ offers additional potential sensitivity at still greater computational cost.

In practice, for a computationally bound analysis, such as an all-sky search for unknown sources in binary systems, it may be attractive to use only the aligned H1-L1 pair with marginalization over the full $\cos\iota$ range, in order to permit searching over a broader volume of source parameter space. Including V1 into the network may prove beneficial for follow-up studies of interesting outlier candidates from an H1-L1 analysis.

\subsection{\label{sec:efficiencyCurves}Detection efficiency curves}
A useful measure of improvement of the TwoSpect sensitivity is the detection efficiency as a function of strain amplitude, $h_0$, for a fixed false alarm rate. To study this, a Monte Carlo simulation of different waveforms using LALSuite software~\citep{LALrepository} is generated for different interferometer combinations: H1-L1, H1-V1, L1-V1, H1-L1-V1, as well as for the single interferometers for each simulated waveform.

Each simulated source has a unique random instantiation of Gaussian noise for each interferometer with equal-amplitude spectral noise density of $S_h^{1/2}=10^{-23}$~Hz$^{-1/2}$. The waveforms are uniformly selected over the sky, with a frequency, $f$, randomly chosen uniformly to lie within 200~Hz to 200.25~Hz, the neutron star is randomly oriented (uniform in $\cos\iota$ and $\psi$), in a circular orbit with a companion with a period of modulation randomly chosen in the range of $2\,{\rm{h}}\leq P\leq 2,252.85\,{\rm{h}}$ and a period-dependent amplitude of frequency modulation spanning $0.278\,{\rm{mHz}}\leq\Delta f\leq 100\,{\rm{mHz}}$ (such that the signal remains in a single frequency bin during each SFT). These parameters are the same search space as for~\cite{TwoSpectS6VSR23}. Different $h_0$ values were randomly selected from the logarithmic space from $5\times10^{-26}$ to $5\times10^{-24}$.

For each of these TwoSpect simulations, a detection for a given source is claimed if there is at least one outlier remaining at the end of the pipeline with a detection statistic that exceeds a threshold set by a false alarm probability (FAP). The TwoSpect pipeline searches the parameters $(f,P,\Delta f)$ with a multi-stage pipeline; the first stage, incoherent harmonic sum detection statistic false alarm threshold is set at $\textrm{FAP}=10^{-10}$, and the second stage, template detection statistic, $R$, false alarm threshold is set at $\textrm{FAP}=10^{-14}$. For further details regarding TwoSpect detection statistics and false alarm probability see~\citep{TwoSpectMethod,TwoSpectS6VSR23}.

\begin{figure}
\centering
\includegraphics[width=0.9\textwidth]{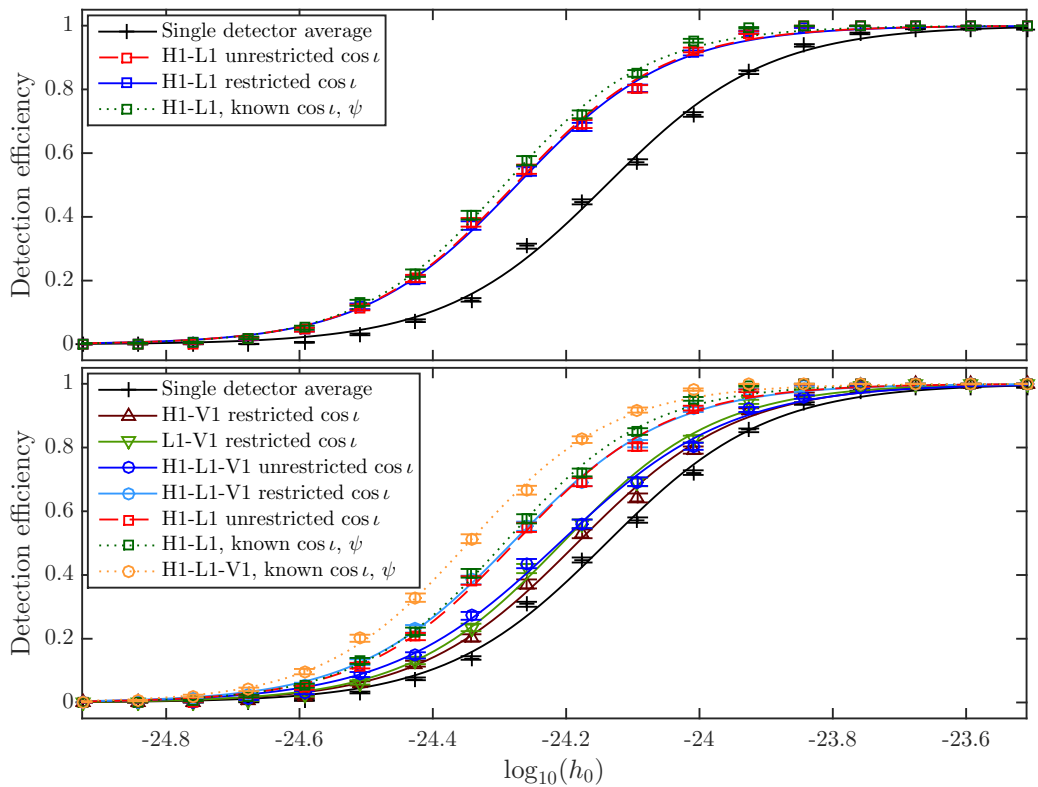}
\caption{\label{fig:efficiency}Efficiency curves for a selected false alarm probability of the TwoSpect pipeline (the first stage incoherent harmonic sum, $10^{-10}$; and second stage template $10^{-14}$~\citep{TwoSpectMethod}) for different coherent SFT sums of interferometer data or single detector analyses. Solid and dashed lines indicate where source orientation for the simulated signals is unknown. Dotted lines indicate where the values of $\cos\iota$ and $\psi$ are known and thus yielding additional improvements in strain sensitivity. Error bars show the 68\% uncertainty region of each set of injections around histogrammed values of $\log_{10}(h_0)$, and the curves are least-square fits to the data points. Top plot: average single-detector analysis compared with 3 different scenarios of H1-L1 coherent analyses. Bottom plot: comparing different coherent analyses that include V1. Observe that the H1-L1-V1 combination where $\cos\iota$ and $\psi$ are known performs best, whereas H1-L1-V1 restricted and H1-L1 unrestricted perform nearly identically when $\cos\iota$ and $\psi$ are unknown, and H1-L1-V1 unrestricted performs poorly compared to the restricted case.}
\end{figure}

As shown in figure~\ref{fig:efficiency} (bottom plot) and summarized in table~\ref{tab:efficiency}, the most sensitive combination is H1-L1-V1 when the polarization parameters $\cos\iota$ and $\psi$ are known. At a detection efficiency of 90\%, the corresponding strain sensitivity for the H1-L1-V1 combination with known polarization parameters is $h_0\approx7.8\times10^{-25}$. On the other hand, the average single detector strain sensitivity is $h_0\approx1.34\times10^{-24}$. This means the improvement over a single-detector analysis is $\approx$42\%. For the H1-L1 coherent sum with known values of $\cos\iota$ and $\psi$, the strain sensitivity is $h_0\approx8.8\times10^{-25}$, or an improvement of $\approx$34\% over the average single detector. 

In the case where polarization parameters are unknown, the strain sensitivity of the H1-L1-V1 combination with restricted $\cos\iota$ marginalization is $h_0\approx9.5\times10^{-25}$, or an improvement of $\approx$29\% over the single-detector analysis, while the H1-L1 unrestricted analysis yields a slightly better strain sensitivity of $h_0\approx9.4\times10^{-25}$ (see figure~\ref{fig:efficiency}, bottom). Alternatively, marginalizing over the full range of $\cos\iota$ in the H1-L1-V1 case, would have improved the sensitivity by only $\approx$10\% over the average single-detector analysis. Analyses that use this restricted $\cos\iota$ marginalization, however, require roughly twice the nominal computational cost for the same set of detectors. In fact, the H1-L1-V1 restricted $\cos\iota$ analysis requires 4 times the number of SFT correction factors to be computed compared to the H1-L1 unrestricted $\cos\iota$ analysis, which is a major computational burden and does not improve the all-sky analysis (see figure~\ref{fig:efficiency} and table~\ref{tab:efficiency}).

When $\cos\iota$ and $\psi$ are unknown, including V1 in any coherent analysis with either LIGO interferometer requires marginalizing $\cos\iota$ over a subset of values in order to improve substantially over a single-detector analysis (see figure~\ref{fig:efficiency}, bottom, and table~\ref{tab:efficiency}). The H1-L1 pair, however, can improve over the single-detector analysis even with marginalization over the full range of $\cos\iota$ values. Importantly, in this case, the H1-L1 sensitivity is slightly better than the H1-L1-V1 coherent combination and has significantly reduced computational burdens. In the H1-L1-V1 analysis, V1 contributes only marginally to enhancing the sensitivity when $\cos\iota$ or $\psi$ are unknown. This effect is also observed when measuring detection efficiency of H1-V1 or L1-V1 searches. The H1 (or L1) and V1 detectors are sensitive to different sky regions and different gravitational wave polarizations so that, when coherently summing SFTs, one or the other detector contributes most of the signal.
\Table{\label{tab:efficiency}Summary of different all-sky TwoSpect search analysis detection sensitivities at 90\% efficiency, where each interferometer strain noise is $S_h^{1/2}=10^{-23}$~Hz$^{-1/2}$, and improvements with respect to the average single-detector analysis.}
\br
Analysis type & $h_0^{90\%}$ sensitivity & Improvement \\
& $(\times 10^{-25})$ & \\
\mr
Average single detector & 13.4 & \0\0- \\
H1-V1 restricted $\cos\iota$ & 12.0 & 10\% \\
H1-L1-V1 unrestricted $\cos\iota$ & 11.8 & 12\% \\
L1-V1 restricted $\cos\iota$ & 11.2 & 16\% \\
H1-L1-V1 restricted $\cos\iota$ & \09.5 & 29\% \\
H1-L1 restricted $\cos\iota$ & \09.5 & 29\% \\
H1-L1 unrestricted $\cos\iota$ & \09.4 & 30\% \\
H1-L1 known $\cos\iota$, $\psi$ & \08.8 & 34\% \\
H1-L1-V1 known $\cos\iota$, $\psi$ & \07.8 & 42\% \\
\br
\endTable

\subsection{All-sky upper limits}
A second useful measure of improvement of the TwoSpect sensitivity is the reduction of the all-sky upper limits derived from a single detector to those derived from the coherent combination of data from two or more detectors. TwoSpect determines the all-sky upper limits on $h_0$ from the loudest outlier statistic of the first stage of the analysis, incoherent harmonic summing~\citep{TwoSpectMethod}. The 95\% confidence level upper limit on $h_0$ is computed by assuming the loudest outlier is due to noise alone, using a root-finding algorithm to determine a $\chi^2$ non-centrality parameter that would cause the detection statistic to exceed the outlier 95\% of the time, then turning this new value of the detection statistic into an $h_0$ value. The scale factor to convert this value is determined empirically using injections, finding the value that correctly sets the upper limit above the injected value 95\% of the time. This procedure is described in further detail in~\citep{TwoSpectS6VSR23}.

Again, a Monte Carlo simulation of different waveforms is generated for three different cases: a single detector, H1; a coherent combination of H1-L1 (with unrestricted marginalization on $\cos\iota$); and a coherent combination of H1-L1-V1 (with restricted marginalization on $\cos\iota$). Other single-detector upper limits (i.e. L1 and V1) do not need to be computed because only the most stringent upper limit (the lowest) of two or more detectors is used, and a similar simulation for other detectors yields consistent results when averaged over the sky.

\begin{figure}
\centering
\includegraphics[width=0.95\textwidth]{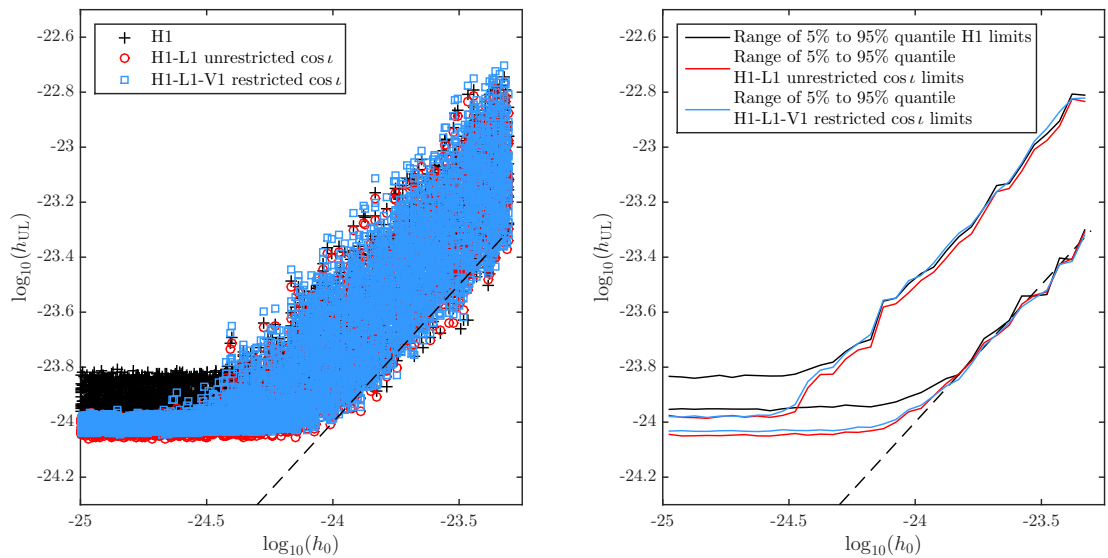}
\caption{\label{fig:ulValidation}Simulated TwoSpect upper limits for randomly polarized gravitational waves; the left- and right-hand plots are of the same data, with the right-hand plot showing the range of values when the injections are binned. For these tests, 3 different analysis cases are explored: 1) A single-detector analysis (black crosses/solid black lines), 2) a coherent sum of H1-L1 SFTs (red circles/solid red lines), and 3) a coherent sum of H1-L1-V1 SFTs (blue squares/solid blue lines). The sensitivity, defined by the upper limit in the low signal regime (left side of the plots), is improved by $\approx$31\% for the H1-L1 analysis and $\approx$29\% for the H1-L1-V1 analysis over the H1 analysis alone.}
\end{figure}

The simulations are over the same parameters as given in section~\ref{sec:efficiencyCurves}. As shown in figure~\ref{fig:ulValidation}, the upper limits for the coherent SFT analyses are lower than the single-detector analysis. The region where $h_0 \rightarrow 0$ defines another measure of search sensitivity. In this region, the H1-L1 coherent analysis yields an improvement of $\approx$31\% over the single-detector analysis and the H1-L1-V1 analysis also yields an improvement of $\approx$29\%. These improvements in upper limits are consistent with the improvements in strain sensitivity at 95\% detection efficiency, and within the range of uncertainties of the sigmoid fits of the H1-L1 unrestricted and H1-L1-V1 restricted analyses described in section~\ref{sec:efficiencyCurves}.

\section{Conclusions}
We have presented a method to sum SFTs coherently from multiple detectors, and have shown resulting improvements to TwoSpect all-sky binary detection efficiency and upper limits computed for non-detections. The best-case strain sensitivity improvement for TwoSpect is $\approx$42\% when a search is able to exploit knowledge of $\cos\iota$ and $\psi$, such as in a search for continuous gravitational waves from Scorpius X-1 under the assumption that the stellar spin aligns with observed radio jets. A typical all-sky search, however, will not determine these parameters; at least, not in the first stage of analysis. In the typical all-sky case, the strain sensitivity is improved by $\approx$30\% for randomly oriented signals. The all-sky upper limit improvement is $\approx$31\% for the H1-L1 pair, and $\approx$29\% for the H1-L1-V1 coherent SFT combination. The expected improvement for unequal strain noise detectors would be less than the values presented here, although still better than analyzing data from detectors separately. Potential sensitivity gains are also reduced by observation periods when detectors are not operating simultaneously.

In order to make an all-sky search computationally tractable, some assumptions must be made of the signal. This analysis reveals that mis-aligned detectors yield limited improvement in detection efficiency when coherently summing SFTs using marginalization over the full range of $\cos\iota$ and $\psi$. In order to recover such a loss, a more restricted marginalization must be made over positive and negative $\cos\iota$ values separately, at significant additional computational cost.

So-called ``directed searches''---analyses that are able to exploit knowledge of source parameters---could potentially search over the unknown polarization parameters and further improve strain sensitivity and upper limits for non-detections. Additionally, the coherent sum could be further improved with an assumption of the gravitational wave frequency. The computational costs must be carefully considered, however. We anticipate that candidates that survive the all-sky analysis pipeline with reasonably small uncertainties on the other observed signal parameters can be followed-up with a search over $\cos\iota$ and $\psi$, to either improve the signal-to-noise ratio of a candidate signal or to reject these outliers as detector artifacts.

\ack
We thank B. Krishnan, G. Meadors, J. Whelan, and the continuous waves working group of the LIGO Scientific Collaboration and Virgo Collaboration for helpful discussions. The authors are grateful for the support of the Division of Observational Relativity and Cosmology of the Albert Einstein Institute Hannover, Max Planck Society, and the National Science Foundation for funding (grant PHY-1205173). This article has document number LIGO-P1500038.

\bibliography{references}

\end{document}